\documentclass[10pt,a4paper]{article}

\usepackage[utf8]{inputenc}
\usepackage{amsmath,amssymb}
\usepackage{microtype}
\usepackage[scaled=0.92]{helvet}
\newcommand{\interfont}{\fontencoding{T1}\fontfamily{phv}\selectfont}
\newcommand{\interdisplay}{\fontencoding{T1}\fontfamily{phv}\fontseries{b}\selectfont}

\usepackage[a4paper,top=2.5cm,bottom=2.5cm,left=2.5cm,right=2.5cm]{geometry}
\usepackage[dvipsnames]{xcolor}
\definecolor{linkblue}{RGB}{6,104,225}
\definecolor{headgray}{RGB}{28,30,33}
\definecolor{subtle}{RGB}{90,92,96}
\definecolor{boxgray}{RGB}{241,243,245}
\usepackage[most]{tcolorbox}

\usepackage{graphicx}
\graphicspath{{figures/}}
\usepackage{booktabs}
\usepackage{array}
\usepackage{tabularx}
\usepackage{float}

\usepackage{caption}
\DeclareCaptionFont{interbold}{\interfont\bfseries}
\DeclareCaptionLabelSeparator{wide}{\hspace{0.6em}}
\usepackage{titlesec}
\titleformat{\section}{\interfont\fontsize{13}{16}\selectfont\bfseries\color{headgray}}{\thesection}{0.8em}{}
\titleformat{\subsection}{\interfont\fontsize{11}{14}\selectfont\bfseries\color{headgray}}{\thesubsection}{0.8em}{}
\titleformat{\subsubsection}{\interfont\normalsize\bfseries\color{headgray}}{\thesubsubsection}{0.8em}{}
\titlespacing*{\section}{0pt}{1.4em}{0.5em}
\titlespacing*{\subsection}{0pt}{1.0em}{0.3em}
\titlespacing*{\subsubsection}{0pt}{0.8em}{0.2em}

\usepackage[numbers,sort&compress,square]{natbib}
\setcitestyle{citesep={, }}
\usepackage{url}
\usepackage[colorlinks=true,linkcolor=linkblue,citecolor=linkblue,urlcolor=linkblue]{hyperref}

\newcommand{\ds}[1]{\texttt{#1}}      
\newcommand{\metaline}[2]{\noindent{\interfont\small\textbf{#1:}\enspace #2}\par}

\begin{document}

\thispagestyle{plain}
\vspace*{-1.2em}
\begin{tcolorbox}[enhanced,colback=boxgray,colframe=boxgray,boxrule=0pt,arc=4mm,
  left=6mm,right=6mm,top=5mm,bottom=5mm,width=\linewidth]
\setlength{\parskip}{0pt}
{\raggedright\hyphenpenalty=10000\interdisplay\fontsize{18}{22}\selectfont\color{headgray}
Learning the Past from the Future: Retrospective Reconstruction in Temporal Interaction Networks\par}
\vspace{0.8em}
{\interfont\small\bfseries Amarnath R\textsuperscript{\normalfont 1}\par}
\vspace{0.6em}
{\small\textsuperscript{1}Department of Information Technology, United Institute of Technology, Coimbatore, India\par}
\vspace{1.1em}
{\setlength{\parskip}{0pt}
Temporal link prediction uses past interactions to predict future links. In practice, interaction records from web and mobile applications can contain gaps caused by logging failures, API limits or lost records, even when later events remain available. To our knowledge, the extent to which later events help recover interactions within such gaps has not been measured directly for this ranking task. We study retrospective reconstruction using the \ds{tgbl-wiki} and \ds{tgbl-comment} datasets. We hide a two-hour window, use its interactions as ground truth, and rank the directed pairs active within it. We compare count-based heuristics and logistic regression models using past information alone or together with 1, 6 or 24 hours of later events. Target pairs and candidate lists remain fixed across conditions, so differences in ranking performance cannot be attributed to changes in candidate selection. With 24 hours of later events, the 26-feature logistic regression improves mean reciprocal rank (MRR) by 0.039 on \ds{tgbl-wiki} under exhaustive ranking, 0.046 under historical-random ranking, and 0.108 on \ds{tgbl-comment} under historical-random ranking. The 26-feature logistic regression benefits from adding later features to past features, whereas replacing the past pair count with the later pair count reduces MRR by 0.157 on \ds{tgbl-wiki} under exhaustive ranking. A control removing later interactions from query sources to historically observed destinations reduces the MRR gain by 59\% on \ds{tgbl-wiki} and 10\% on \ds{tgbl-comment}. Because this control removes more than repeated target pairs, these reductions cannot be attributed to pair recurrence alone. Later events improve the ranking of hidden interactions on both datasets, including \ds{tgbl-comment}, where only 6.4\% of target pairs reappeared within 24 hours.
\par}
\vspace{1.2em}
\metaline{Keywords}{retrospective reconstruction, later events, directed interaction pairs, logistic regression, fixed candidate lists, temporal interaction networks}
\vspace{0.25em}
\metaline{Date}{October 23, 2026}
\vspace{0.25em}
\metaline{Correspondence}{\href{mailto:amarnathresearch@gmail.com}{amarnathresearch@gmail.com}}
\vspace{0.25em}
\metaline{ORCID}{\href{https://orcid.org/0000-0003-3513-8969}{0000-0003-3513-8969}}
\end{tcolorbox}
\vspace{0.4em}

\section{Introduction}

Temporal link prediction uses past events to predict future links, assuming that the recorded history is mostly available~\cite{poursafaei2022,qin2023,huang2023tgb}. In practice, interaction records from web and mobile applications, such as Wikipedia, Reddit, Twitter/X and messaging apps, may be incomplete because of logging failures, API limits, crawler errors, or lost or deleted records. As a result, some interactions may be missing even when records from before and after the affected period remain available. These gaps can affect the analysis of network activity and the reliability of predictions based on interaction history. When later records are available, they may provide useful information about interactions from the missing period. However, conventional temporal link prediction evaluates how well past observations predict future links, not whether later events can help reconstruct earlier interactions.

To our knowledge, previous work has not directly evaluated how much events recorded after a missing interval improve the ranking of directed interaction pairs that were active during that interval, with the target pairs and candidate lists held fixed across information conditions. Previous studies have addressed missing links and events using temporal patterns and network structure~\cite{bagrow2021,stomakhin2011,guimera2009,rodrigues2025}, and temporal knowledge graph completion has used information from both before and after a query time to infer missing facts~\cite{cai2023,chen2024,goel2020}, but we did not find this specific comparison reported in the studies reviewed here. This study measures that improvement using graph-derived features, count-based heuristics and logistic regression models, while keeping target pairs and candidate lists fixed across conditions, so that differences in ranking performance are not caused by changes in candidate selection.

Evaluating retrospective reconstruction requires a dataset in which the interactions hidden during the experiment remain available as ground truth. We therefore conduct experiments on two temporal interaction datasets from the Temporal Graph Benchmark~\cite{huang2023tgb}, \ds{tgbl-wiki} and \ds{tgbl-comment}, which contain Wikipedia edit events and Reddit reply events, respectively. In each dataset, we intentionally hide a two-hour window and rank the directed pairs that were active during that window. The two datasets differ in how often these pairs appear again, as 43.0\% of them reappear within 24 hours in \ds{tgbl-wiki} but only 6.4\% in \ds{tgbl-comment}. The models use either the history available before the window or that history together with the events from the following 1, 6 or 24 hours. We also run removal and destination-shuffling controls to examine how changes to the later events affect the results.

This study makes three contributions. First, we set up retrospective reconstruction as a ranking task with fixed candidate lists. Second, we show that with 24 hours of later events, mean reciprocal rank (MRR) improves under every candidate protocol tested. The gains range from 0.022 to 0.046 on \ds{tgbl-wiki} across degree-matched, exhaustive and historical-random ranking, and from 0.108 to 0.134 on \ds{tgbl-comment} across historical-random and degree-matched ranking. Third, we find that removing later interactions from query sources to historically observed destinations reduces the historical-random gain by 59\% on \ds{tgbl-wiki} but only by 10\% on \ds{tgbl-comment}. Because this control also removes interactions other than repeated target pairs, these reductions cannot be attributed to repeated pairs alone.

\section{Related Work}

\subsection{Temporal Link Prediction}
Temporal link prediction uses observed interaction histories to predict links that occur in the future~\cite{qin2023}. Poursafaei et al.~\cite{poursafaei2022} showed that evaluation results depend on how negative samples are selected and introduced EdgeBank, a baseline that predicts edges observed previously. The Temporal Graph Benchmark~\cite{huang2023tgb} provides datasets and evaluation protocols for comparing temporal graph models. Recent work has also examined the contribution of simple signals. Qarkaxhija and Scholtes~\cite{qarkaxhija2026} proposed a smoothed-count baseline with a small number of parameters, while Zhang et al.~\cite{zhang2026} argued that common negative-sampling protocols hide source-node activity from link predictors. Following these studies, we include a pair-count heuristic related to EdgeBank and a six-feature logistic regression model based on node activity. These studies aim at stronger forward predictors and evaluate links in the forward direction. They do not measure whether events recorded after a gap help rank pairs that were active during it, which is the question we address with count-based and activity features under fixed candidates.

\subsection{Network Reconstruction}
Network reconstruction methods infer missing or unreliable information from observed network structure and activity. Guimer\`a and Sales-Pardo~\cite{guimera2009} developed a framework for identifying missing and spurious links in static networks. Stomakhin, Short and Bertozzi~\cite{stomakhin2011} studied the recovery of missing participants in social-network events from interaction timing. Bagrow and Lehmann~\cite{bagrow2021} investigated the recovery of edge-level activity from node-level activity and network structure, while Rodrigues~\cite{rodrigues2025} reviewed prediction and inference methods for complex networks. These studies address different forms of missing network information under specific assumptions. The present study instead examines whether events recorded after a gap improve the ranking of directed pairs from that period.

\subsection{Temporal Knowledge Graph Completion}
Temporal knowledge graph completion infers missing facts using information associated with time. It distinguishes interpolation, which can use facts from before and after the query time, from extrapolation, which predicts facts beyond the observed time range~\cite{cai2023}. Goel et al.~\cite{goel2020} introduced diachronic entity embeddings for temporal knowledge graph completion, while Chen et al.~\cite{chen2024} proposed a model for both interpolation and extrapolation. This literature establishes that using information from after a query time to infer earlier facts is not, by itself, a new idea. Our study differs in its prediction target. Rather than ranking missing entities or facts in a multi-relational knowledge graph, we rank directed interaction pairs that were active at least once during a two-hour window in a temporal interaction network.

\section{Methods}

\subsection{Data and Episode Construction}
We use two datasets from the Temporal Graph Benchmark (TGB)~\cite{huang2023tgb}, \ds{tgbl-wiki} and \ds{tgbl-comment}.\linebreak[1] Both datasets are temporal graphs represented as streams of timestamped directed edges $(u, v, t)$, where $u$ is the source, $v$ is the destination and $t$ is the event timestamp. The \ds{tgbl-wiki} dataset contains 157,474 events in which users edit Wikipedia pages. The \ds{tgbl-comment} dataset contains 44.3 million directed reply events among 994,790 Reddit users. We remove the 1,333,961 self-loop events in \ds{tgbl-comment} but retain repeated interactions between the same source and destination, since repeated events may provide useful information for retrospective reconstruction. All timestamps are expressed in seconds.

We place the validation and test boundaries at the 70th and 85th percentiles of the raw timestamps, calculated before removing self-loops. This preserves alignment with the benchmark masks. Each dataset is divided into 12 non-overlapping episodes, comprising six training episodes, three validation episodes and three test episodes.

\begin{figure}[!htbp]
  \centering
  \includegraphics[width=\linewidth]{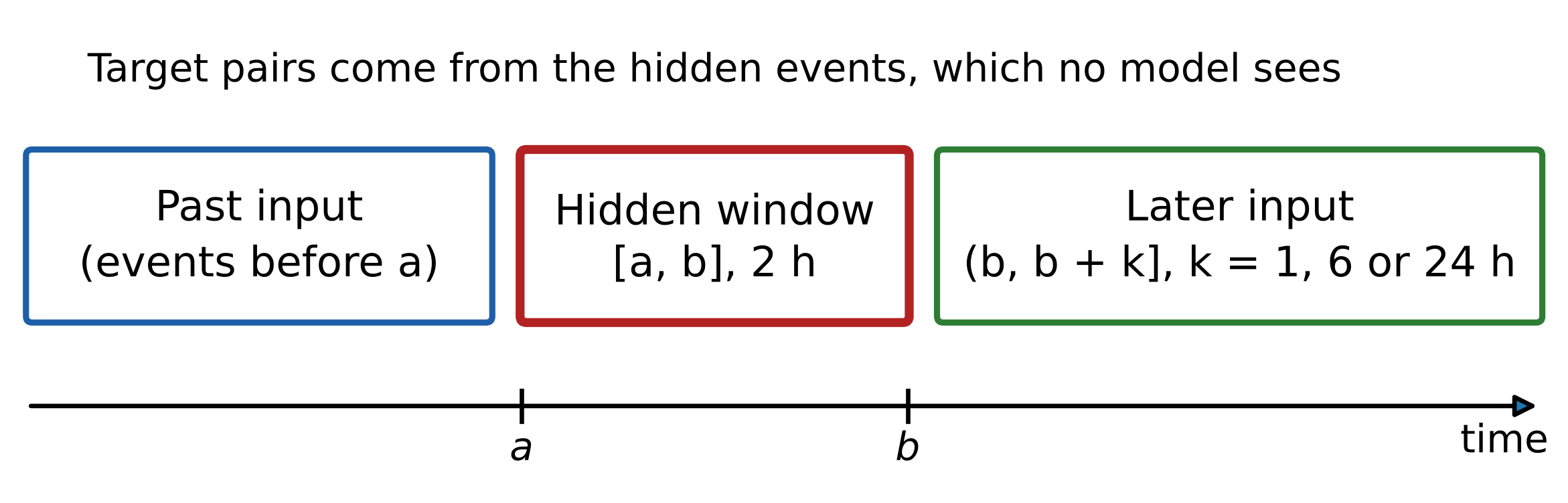}
  \caption{Episode construction. The hidden window is excluded from all model inputs, and target pairs are identified from the events within that window.}
  \label{fig:episode}
\end{figure}

\begin{figure}[!htbp]
  \centering
  \includegraphics[width=\linewidth]{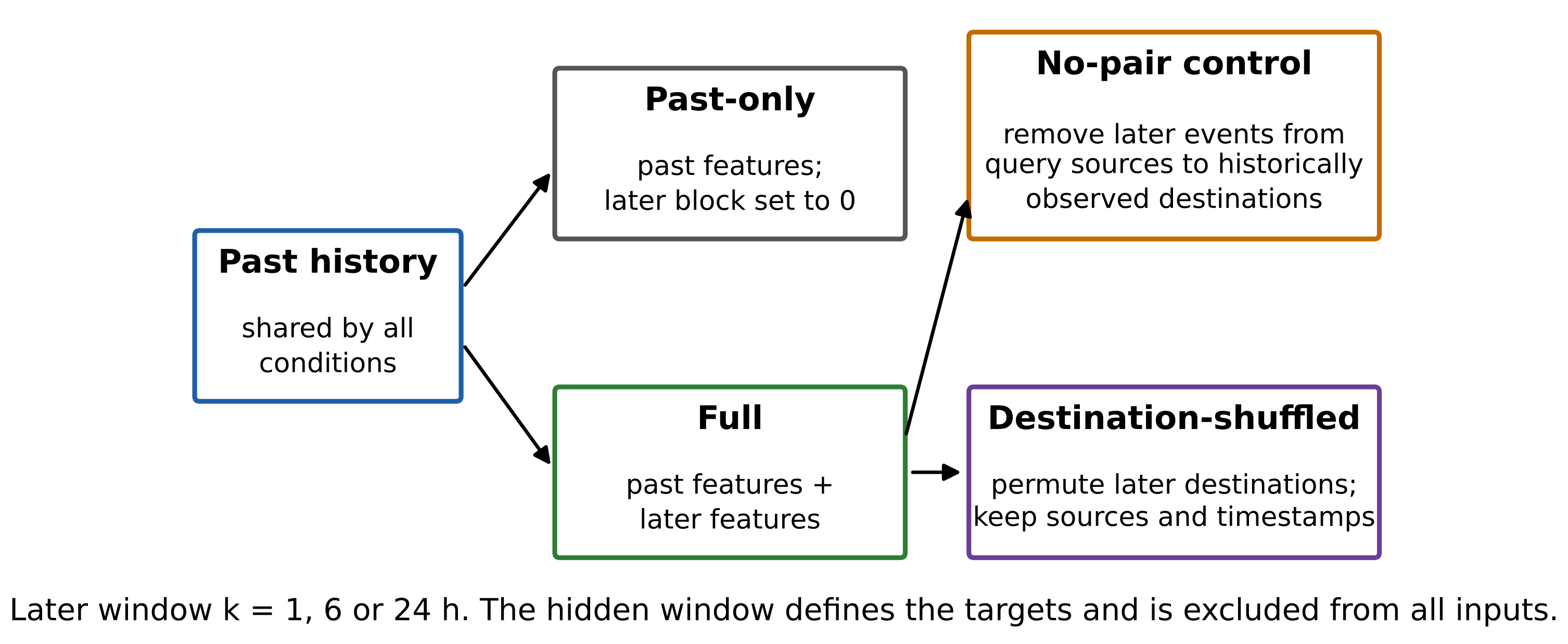}
  \caption{Information conditions and controls. All conditions share the same past history and hidden-window targets.}
  \label{fig:conditions}
\end{figure}

Each episode contains a hidden window $[a, b]$ lasting two hours. For \ds{tgbl-wiki}, the past input includes all events recorded before $a$. For \ds{tgbl-comment}, it includes only events from the 30 days before $a$, because processing the full history for every episode is computationally impractical. The later input contains events recorded in $(b, b + k]$, where $k$ is 1, 6 or 24 hours. Events inside the hidden window are excluded from all model inputs and are retained only to identify the target pairs. Figure~\ref{fig:episode} illustrates the episode construction, and Figure~\ref{fig:conditions} shows the information conditions.

\subsection{Targets and Candidate Construction}
A target is a distinct directed pair $(u, v)$ with at least one event in the hidden window. Multiple events involving the same pair count as a single target because the task is to rank pairs that were active during the window, rather than estimate the number of interactions. We retain a target only if its destination $v$ has appeared in the past input. These destinations define the historical universe, restricting evaluation to destinations that a model using only past information could rank.

We cap the number of targets at 300 per episode for \ds{tgbl-wiki} and 1,000 per episode for \ds{tgbl-comment}. When a cap applies, targets are sampled uniformly without replacement using one seed per episode, shared across all model seeds. None of the three \ds{tgbl-wiki} test episodes reached the 300-target cap, and they contain 214, 138 and 132 targets, respectively. These episodes produce 484 target lists from 384 source users. For \ds{tgbl-comment}, the cap applies to all three test episodes, producing 3,000 target lists from 2,686 source users.

Each target pair defines a separate ranking list. A source with multiple target destinations therefore contributes multiple lists. Each list contains the true destination and a set of negative candidates. A negative is a destination that the source did not reach during the hidden window, which does not make it an impossible interaction. Candidate lists are constructed using the past input and hidden-window labels, without using any events recorded after $b$. The same lists are used across all information conditions. We evaluate three candidate protocols, summarized in Table~\ref{tab:protocols}.

\begin{table}[!htbp]
  \centering
  \caption{Candidate protocols.}
  \label{tab:protocols}
  \small
  \begin{tabularx}{\linewidth}{@{}l X p{4.3cm}@{}}
    \toprule
    \textbf{Protocol} & \textbf{Construction} & \textbf{Role}\\
    \midrule
    Exhaustive & Every destination in the historical universe & Primary for \ds{tgbl-wiki}, not feasible for \ds{tgbl-comment}\\[3pt]
    Historical-random & 100 negatives: up to 50 past destinations of the query source, with the remainder sampled uniformly & Primary for \ds{tgbl-comment}, matched comparison across datasets\\[3pt]
    Degree-matched & 100 negatives from the same past in-degree bin as the true destination, with four bins fitted on training episodes & Sensitivity analysis\\
    \bottomrule
  \end{tabularx}
\end{table}

For exhaustive ranking on \ds{tgbl-wiki}, the evaluation produces 1,676,134 candidate rows, averaging approximately 925 candidates per list. Historical-random ranking supports a matched comparison across the two datasets. Degree-matched ranking tests whether gains persist when negative candidates have similar destination in-degree to the true destination. In all protocols, negative candidates are sampled without replacement from destinations outside an exclusion set containing the true destination, every other destination reached by the source during the hidden window and the source itself. If a degree bin contains fewer than 100 eligible destinations, all eligible destinations are used without filling the list from other bins. Both sampled protocols use five seeds, and candidate lists are identical across information conditions within each seed.

\subsection{Information Conditions and Controls}
We compare four information conditions. In the past-only condition, models use features calculated from events recorded before the hidden window. The later-feature block is set to zero, preserving the same 26-column design matrix across conditions. In the full condition, models use the past features together with features calculated from events in the later window $(b, b + k]$.

The no-pair control removes every later event from a query source to any destination in the historical universe. A query source is a source with at least one target in the episode. This control removes pair recurrence, meaning the same pair appearing again after the hidden window, as well as other later interactions from the same source to historically observed destinations. It removes 92\% of query-source later events in \ds{tgbl-comment} and between 99.7\% and 100\% in \ds{tgbl-wiki}. The removal does not depend on the sampled candidate lists. Because the control removes more than repeated target pairs, it does not isolate the effect of repeated pairs.

The destination-shuffling control permutes destination identities across later events while preserving sources and timestamps. Each source retains its event count, every event timestamp remains unchanged and each destination retains its total number of later events. However, shuffling changes which sources interact with which destinations, when destinations first appear in the later record and the two-hop structure of that record. Its results therefore reflect changes to destination assignment and these associated structural effects, rather than destination assignment alone.

\subsection{Features and Models}
For each candidate pair, we calculate 14 past features and 12 later features using only the events permitted under the corresponding information condition. The feature blocks include source out-degree, destination in-degree, pair count, a pair-presence flag, reverse-pair count, two two-hop counts and recency terms. The past block also contains two activity flags.

The two two-hop features count paths of the forms $u \rightarrow x \leftarrow w \rightarrow v$, with $w \neq u$, and $u \rightarrow x \rightarrow v$. The first counts other sources that share a destination with $u$ and also reach $v$. The second counts paths from $u$ to $v$ through an intermediate node. Repeated events are collapsed to a single edge when calculating these two features. All count features are transformed using $\log(1 + x)$. Recency is calculated as
\begin{equation}
  \rho(\Delta t) = \frac{1}{1 + \Delta t / \tau}, \qquad \tau = 3600\ \text{s},
  \label{eq:recency}
\end{equation}
where $\Delta t$ is measured backward from $a$ for past features and forward from $b$ for later features. Source-only features are constant within a ranking list and therefore cannot change the ordering of destinations on their own.

We evaluate four model families. The pair-count heuristic ranks candidates by their past pair count in the past-only condition and by their later pair count when later events are available, which is our later-window adaptation of EdgeBank~\cite{poursafaei2022} rather than a baseline defined in the original paper. The popularity heuristic ranks candidates by past destination in-degree and adds later destination in-degree when later events are available. The six-feature logistic regression model uses past source out-degree, past destination in-degree, two past activity flags, later source out-degree and later destination in-degree. The main model is a logistic regression using all 26 features.

Each logistic regression model is fitted on the training episodes using the L-BFGS solver. Features are standardized using statistics calculated from training rows. All training positives are retained, giving 851 positives for \ds{tgbl-wiki} and 6,000 for \ds{tgbl-comment}. Training negatives are independently retained with probability
\begin{equation}
  r = \min\!\left(1,\ \frac{50\,n_{+}}{n_{-}}\right),
  \label{eq:sampling}
\end{equation}
where $n_{+}$ and $n_{-}$ are the numbers of training positives and negatives, respectively. This sampling scheme retains approximately 50 negatives per positive when negative sampling is required. We add $\log(r)$ to the model logit to correct for the change in class odds introduced by negative sampling. Since this is a constant added to every candidate score, it does not change the ranking.

The regularization strength $C$ is selected using validation MRR from the grid $\{0.01, 0.1, 1, 10\}$, with ties resolved in favor of the smallest $C$. We repeat model selection using the wider grid $\{10^{-4}, 10^{-3}, 0.01,$ $0.1, 1, 10, 100, 1000\}$ as a sensitivity analysis. Test labels are not used during fitting or model selection and are accessed only after test scores have been saved.

\subsection{Evaluation Metrics and Statistical Analysis}
Our primary metric is mean reciprocal rank (MRR), calculated within each ranking list. Lists without negative candidates are excluded. We also report Hits@1, Hits@5 and Hits@10. Pooled ROC-AUC and average precision are secondary metrics because they aggregate candidate rows across lists and measure a different aspect of performance from within-list ranking.

The heuristic models can produce tied scores. We therefore calculate their expected reciprocal rank under uniform random tie-breaking. Let $g$ be the number of negative candidates scored strictly above the true destination and $t$ the number of negative candidates tied with it, excluding the true destination itself. The expected reciprocal rank is
\begin{equation}
  \mathbb{E}[\mathrm{RR}] = \frac{\psi(g + t + 2) - \psi(g + 1)}{t + 1},
  \label{eq:err}
\end{equation}
where $\psi$ is the digamma function.

Our primary contrast is the difference between MRR with past-plus-later information and MRR with past-only information at each observation horizon. We also compare the full condition with each of the two controls at every observation horizon.

We estimate uncertainty using 2,000 paired cluster-bootstrap replicates over test source users. Each sampled user contributes all their ranking lists across the test episodes, and the compared conditions use the same bootstrap draws. Within each replicate, results are averaged over the five seeds. The resulting intervals describe variation across users within the three test episodes and five model fits. They do not represent uncertainty across independently sampled time periods. We do not apply a multiple-comparison correction.

We also split the primary contrast by whether the target pair appeared in the past, whether the same pair appeared again after the hidden window, source and destination degree bins, and test episode. Because the groups based on repeated pairs are defined using events observed after the hidden window, these analyses are descriptive and are not interpreted as causal evidence.

\subsection{Protocol and Amendments}
Before fitting models to the benchmark data, we documented the primary comparison, evaluation metric, controls, subgroup analyses and bootstrap procedure in a protocol document. One control was revised during the study. The initial no-pair implementation removed only later events on candidate pairs, making the removed set dependent on the sampled negatives. After examining the \ds{tgbl-wiki} results, but before any \ds{tgbl-comment} results were available, we changed the control to remove all later events from query sources to historically observed destinations. Under exhaustive ranking on \ds{tgbl-wiki}, both implementations remove the same events, so the \ds{tgbl-wiki} results were unchanged.

Four reporting choices were also fixed after an initial descriptive results table and before the uncertainty intervals were computed. These were averaging across seeds, reporting all model families, using source-degree bins and applying no multiplicity correction. We report these changes explicitly rather than describing the study as fully preregistered.

The experimental pipeline checks feature-file hashes and candidate-set digests between stages and recalculates evaluation metrics from saved scores. All planned runs using the main regularization grid completed, with 420 runs on \ds{tgbl-wiki} and 280 on \ds{tgbl-comment} and no logistic regression convergence failures. Before running the wider regularization grid, we established a stability criterion requiring every specified contrast to retain both its sign and its interval status, that is, whether its 95\% interval includes zero, at every observation horizon.

\section{Results}

\subsection{Later Events Improve Ranking at Every Horizon}
\begin{figure}[!htbp]
  \centering
  \includegraphics[width=0.85\linewidth]{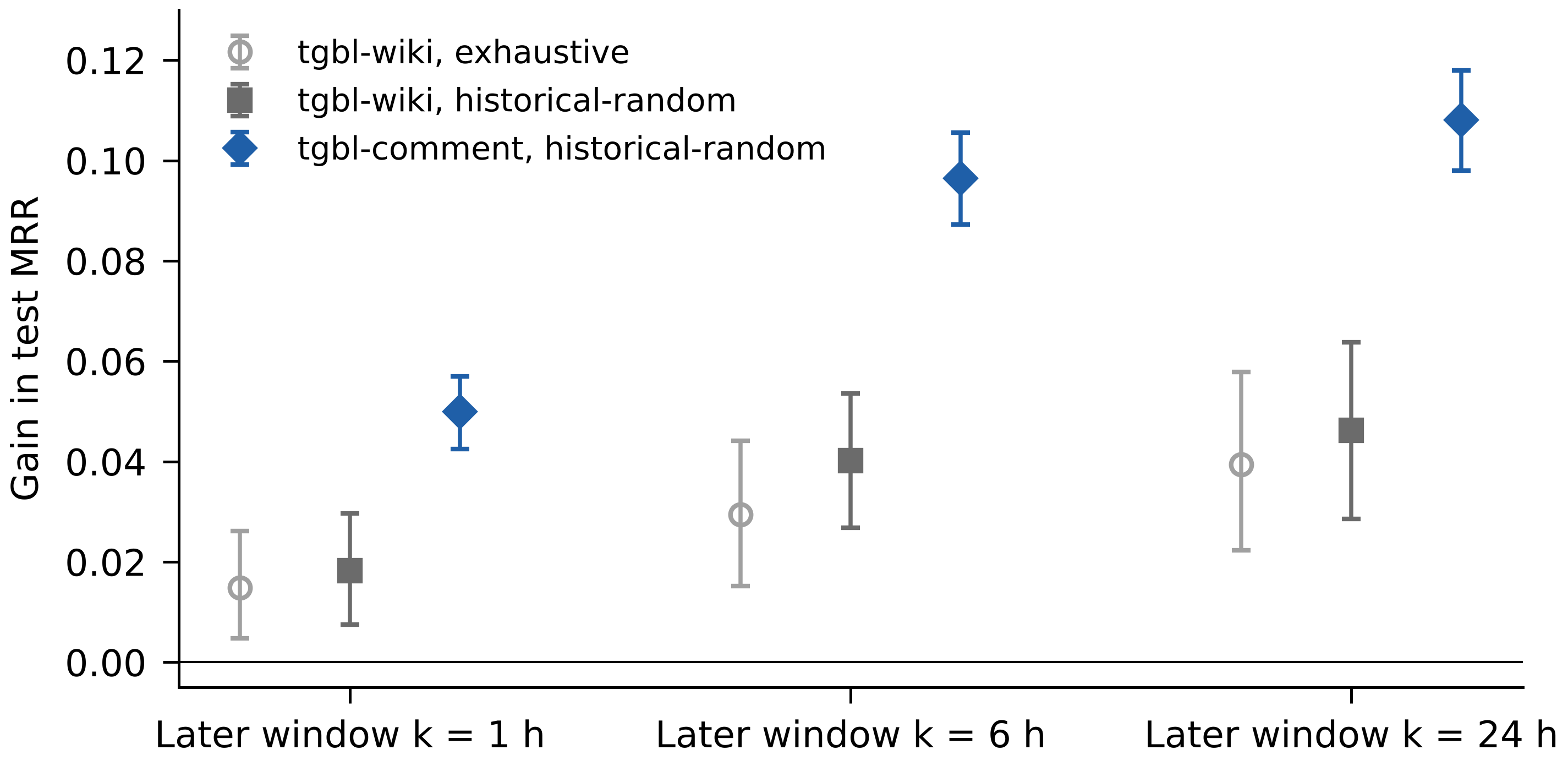}
  \caption{Gain in test MRR from adding later events, for the 26-feature logistic regression at each horizon. Points are seed-averaged differences between the past-plus-later and past-only conditions, and bars are 95\% paired user-cluster bootstrap intervals.}
  \label{fig:horizon}
\end{figure}

Adding later events raised the MRR of the 26-feature logistic regression in all three main comparisons and at all three horizons (Table~\ref{tab:main}, Figure~\ref{fig:horizon}). At 24 hours, MRR rose from 0.599 to 0.639 under exhaustive ranking on \ds{tgbl-wiki}, a gain of 0.039 [0.022, 0.058]. Under historical-random ranking, the gain was 0.046 [0.029, 0.064] on \ds{tgbl-wiki} and 0.108 [0.098, 0.118] on \ds{tgbl-comment}. Every primary interval excluded zero, including those at 1 and 6 hours. As described in Section~3.5, these intervals reflect variation across users within the three test episodes and five model fits, not across independent time periods.

The gain grew with the length of the later window, but at different rates. On \ds{tgbl-comment}, it rose from 0.050 at 1 hour to 0.096 at 6 hours and 0.108 at 24 hours, so 89\% of the 24-hour gain was already present at 6 hours. On \ds{tgbl-wiki} under exhaustive ranking, it rose from 0.015 to 0.029 and then 0.039, so 74\% was present at 6 hours. The direction of the gain held in every test episode, with 24-hour gains of 0.031 to 0.058 on \ds{tgbl-wiki} and 0.090 to 0.132 on \ds{tgbl-comment}.

\begin{table}[!htbp]
  \centering
  \caption{Test MRR of the 26-feature logistic regression with past-only and past-plus-later information. Gains are seed-averaged, and the 24-hour gain is shown with its 95\% paired user-cluster bootstrap interval. Hist.-random denotes historical-random ranking. Gains are computed from unrounded values, so they can differ by 0.001 from the difference of the displayed MRRs.}
  \label{tab:main}
  \footnotesize
  \setlength{\tabcolsep}{3.5pt}
  \begin{tabular}{@{}llccccc@{}}
    \toprule
    \textbf{Dataset} & \textbf{Protocol} & \shortstack{\textbf{Past}\\\textbf{MRR}} & \shortstack{\textbf{Past +}\\\textbf{24\,h}} & \shortstack{\textbf{Gain}\\\textbf{1\,h}} & \shortstack{\textbf{Gain}\\\textbf{6\,h}} & \shortstack{\textbf{Gain, 24\,h}\\\textbf{[95\% interval]}}\\
    \midrule
    \ds{tgbl-wiki}    & exhaustive  & 0.599 & 0.639 & +0.015 & +0.029 & +0.039 [0.022, 0.058]\\
    \ds{tgbl-wiki}    & hist.-random & 0.617 & 0.663 & +0.018 & +0.040 & +0.046 [0.029, 0.064]\\
    \ds{tgbl-comment} & hist.-random & 0.485 & 0.593 & +0.050 & +0.096 & +0.108 [0.098, 0.118]\\
    \bottomrule
  \end{tabular}
\end{table}

\subsection{Adding Later Evidence Versus Replacing Past Evidence}
The 26-feature logistic regression had the highest MRR of the four model families under every protocol (Table~\ref{tab:families}). The popularity heuristic gained only 0.005 to 0.009 at 24 hours. The six-feature logistic regression gained 0.007 under exhaustive ranking on \ds{tgbl-wiki}, but 0.056 and 0.121 under historical-random ranking. These historical-random gains are larger than those of the 26-feature model because the six-feature model starts from a lower past-only MRR, and it still finishes below the 26-feature model. In the past-only condition, the six-feature model ranks exactly like the popularity heuristic, because the only features that vary within a list describe the destination.

The pair heuristic behaved differently. Replacing the past pair count with the later pair count lowered MRR by 0.157 under exhaustive ranking on \ds{tgbl-wiki} and by 0.140 under historical-random ranking, but raised it by 0.064 on \ds{tgbl-comment}. This split is consistent with how often target pairs were already known. On \ds{tgbl-wiki}, 71.7\% of test targets had appeared in the past, while only 43.0\% appeared again within 24 hours, so the swap discarded pair evidence for many targets. On \ds{tgbl-comment}, only 10.7\% of targets had appeared in the past, so the past count had little to lose. The logistic regression keeps the past features and adds the later features, and it gained on both datasets.

\begin{table}[!htbp]
  \centering
  \caption{Change in test MRR at 24 hours from adding later events, by model family.}
  \label{tab:families}
  \footnotesize
  \setlength{\tabcolsep}{4pt}
  \begin{tabular}{@{}>{\raggedright\arraybackslash}p{4.3cm}ccc@{}}
    \toprule
    \textbf{Model family} & \shortstack{\textbf{\ds{tgbl-wiki}}\\\textbf{exhaustive}} & \shortstack{\textbf{\ds{tgbl-wiki}}\\\textbf{hist.-random}} & \shortstack{\textbf{\ds{tgbl-comment}}\\\textbf{hist.-random}}\\
    \midrule
    Pair heuristic (later count replaces past count) & $-0.157$ & $-0.140$ & $+0.064$\\
    Popularity heuristic                             & $+0.006$ & $+0.005$ & $+0.009$\\
    Logistic regression, 6 activity features         & $+0.007$ & $+0.056$ & $+0.121$\\
    Logistic regression, 26 features                 & $+0.039$ & $+0.046$ & $+0.108$\\
    \bottomrule
  \end{tabular}
\end{table}

\subsection{What the Controls Remove}
\begin{figure}[!htbp]
  \centering
  \includegraphics[width=\linewidth]{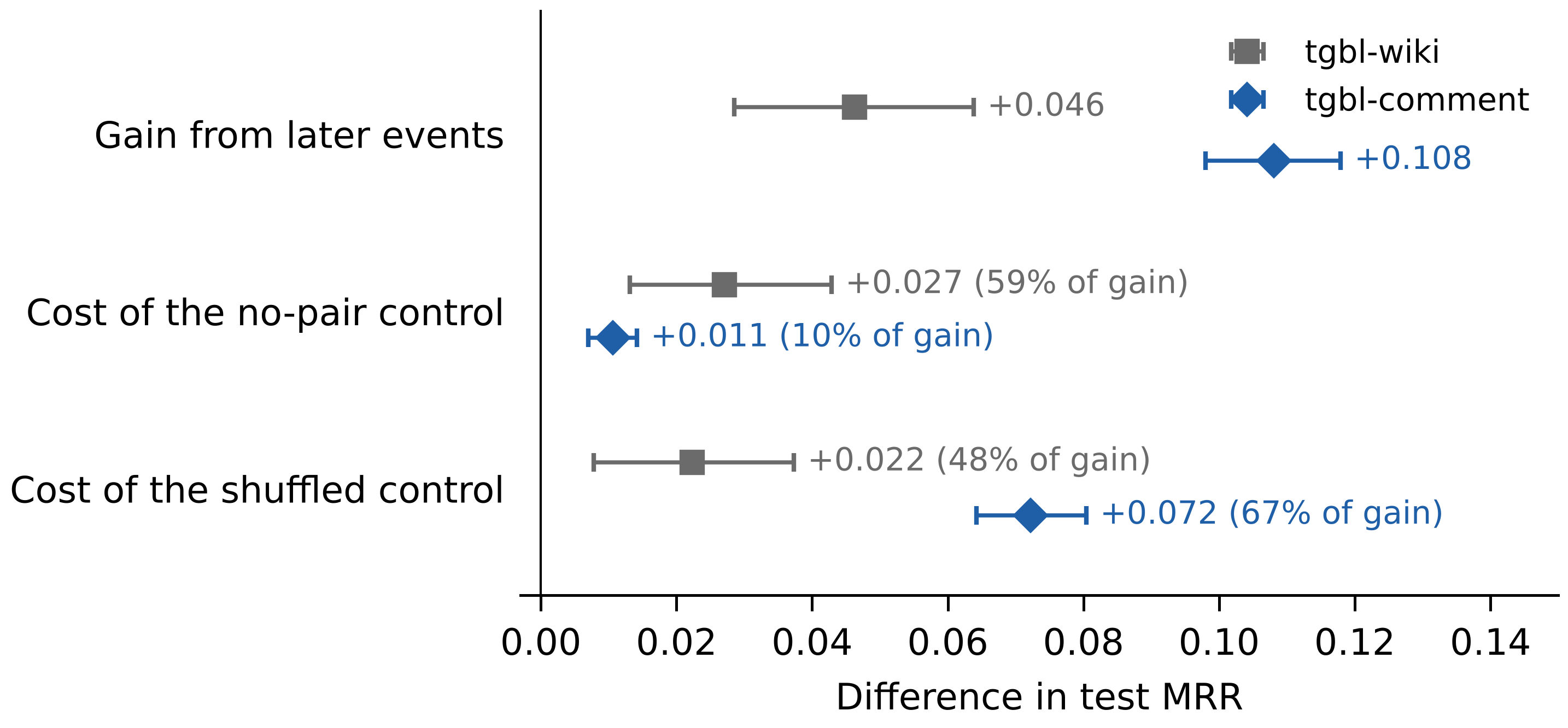}
  \caption{Full gain from later events and the MRR cost of the two controls at 24 hours under historical-random ranking. Bars are 95\% paired user-cluster bootstrap intervals. The shares are ratios of point estimates.}
  \label{fig:controls}
\end{figure}

Under historical-random ranking at 24 hours, the no-pair control cost 0.027 [0.013, 0.043] MRR on \ds{tgbl-wiki}, which is 59\% of the full gain of 0.046 (Figure~\ref{fig:controls}). On \ds{tgbl-comment}, it cost 0.011, or 10\% of the full gain of 0.108. These shares are ratios of point estimates, and we did not compute intervals for them. After the removal, later events still improved MRR over the past-only condition by 0.019 on \ds{tgbl-wiki} and by 0.098 on \ds{tgbl-comment}. These post-removal comparisons were not specified in the original set of primary contrasts. They show that later events outside the removed set still help, and on \ds{tgbl-comment} most of the gain remains after the removal.

Shuffling later destinations cost 0.022 MRR on \ds{tgbl-wiki}, or 48\% of the gain, and 0.072 on \ds{tgbl-comment}, or 67\%. The two controls therefore affect the datasets in opposite ways. On \ds{tgbl-wiki}, the no-pair control reduces the gain more than the shuffle, while on \ds{tgbl-comment} the shuffle reduces it far more than the no-pair control. The no-pair control also removes repeated pairs, and the shuffle changes destination timing and two-hop structure at the same time, so neither control points to a single mechanism. At 1 hour, the shuffled-control interval on \ds{tgbl-wiki} included zero under the main grid (+0.004 [$-0.002$, 0.011]) and excluded it under the wider grid. We therefore treat this contrast as borderline.

\subsection{Gains When Pairs Do Not Appear Again}
\begin{figure}[!htbp]
  \centering
  \includegraphics[width=\linewidth]{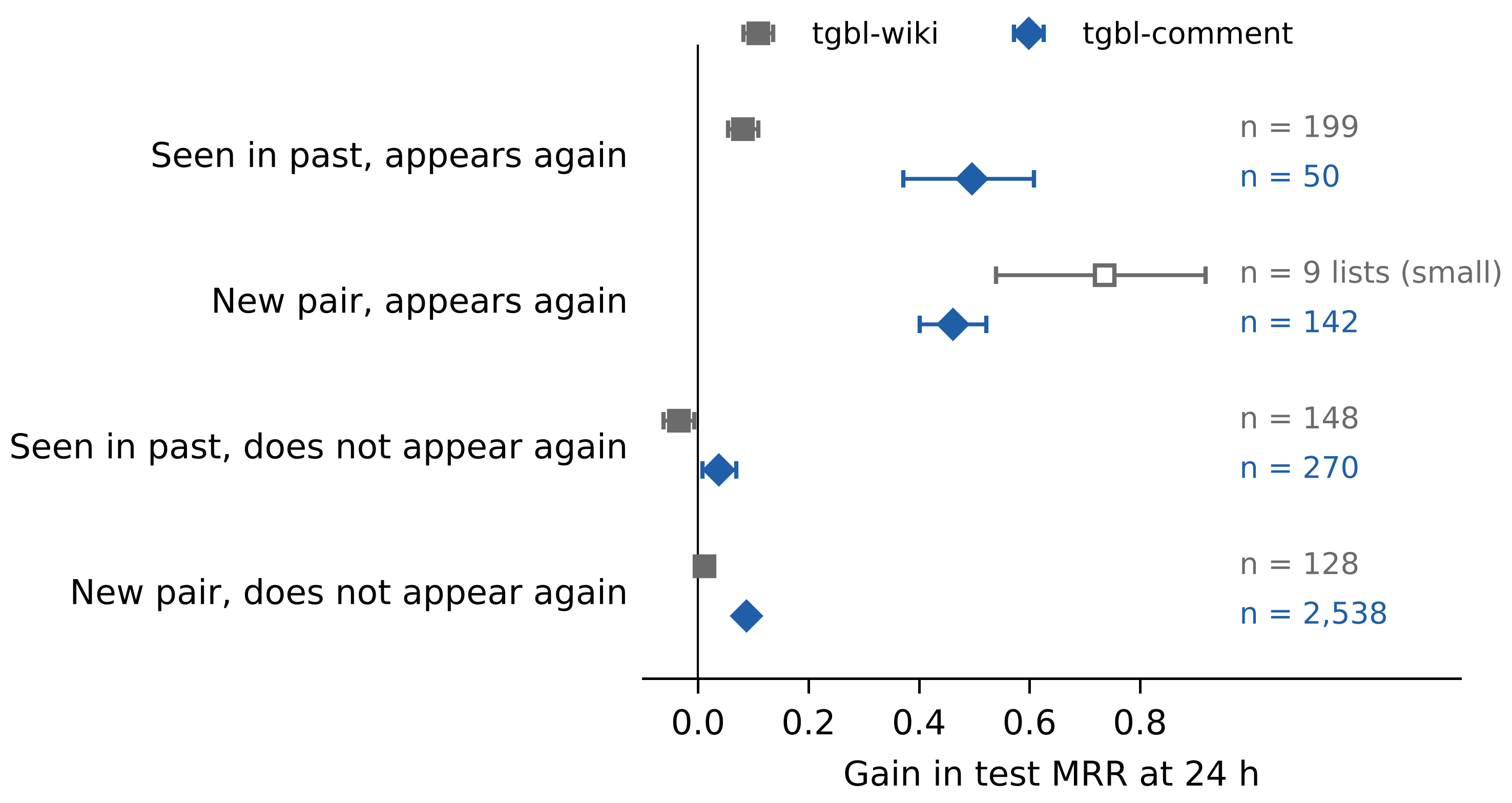}
  \caption{Gain in test MRR at 24 hours by group, under the primary protocol of each dataset. Groups are defined by later outcomes and are descriptive. The \ds{tgbl-wiki} group of new pairs that appeared again holds only 9 lists, and its estimate is highly uncertain.}
  \label{fig:groups}
\end{figure}

Within 24 hours, 43.0\% of test target pairs appeared again on \ds{tgbl-wiki} and 6.4\% on \ds{tgbl-comment}. At 1 and 6 hours, the rates were 14.5\% and 27.7\% on \ds{tgbl-wiki}, and 1.9\% and 3.7\% on \ds{tgbl-comment}. Figure~\ref{fig:groups} splits the 24-hour gain by whether the pair appeared in the past and whether it appeared again later. In both datasets, the largest observed gains were in groups of pairs that appeared again. Pairs seen in the past that appeared again gained 0.081 on \ds{tgbl-wiki} and 0.496 on \ds{tgbl-comment}, and new pairs that appeared again gained 0.461 on \ds{tgbl-comment}. The corresponding \ds{tgbl-wiki} group gained 0.736, but it holds only 9 lists, so this estimate is too unstable to compare with the other groups. The large \ds{tgbl-comment} groups are more informative.

The largest \ds{tgbl-comment} group shows a different pattern. It holds 2,538 of the 3,000 lists, all pairs that neither appeared in the past nor appeared again later, and it still gained 0.088. Weighted by size, this group contributes $2{,}538 \times 0.088 / 3{,}000 = 0.074$ of the overall 0.108, or about 69\%. Most of the \ds{tgbl-comment} gain therefore does not need the pair itself to appear again. On \ds{tgbl-wiki}, pairs seen in the past that did not appear again lost 0.035, while the same group gained 0.038 on \ds{tgbl-comment}. The sign of the gain agreed between the two datasets in 10 of 12 group-by-horizon cells. Because these groups are defined by later outcomes, we treat the patterns as descriptive only.

Two further splits point the same way. On \ds{tgbl-comment}, destinations in the lowest past in-degree bin gained 0.174, against 0.052 in the highest bin. Under degree-matched candidates, where every negative shares the true destination's past in-degree bin, the gain on \ds{tgbl-comment} was 0.072, 0.123 and 0.134 at 1, 6 and 24 hours. On \ds{tgbl-wiki}, it was 0.006 at 1 hour, with an interval that included zero, then 0.019 at 6 hours and 0.022 [0.011, 0.034] at 24 hours. Matching on popularity therefore does not remove the gain, although it does not rule out some role for popularity.

\subsection{Secondary Metrics and Regularization}
The within-list secondary metrics agreed with MRR. At 24 hours, Hits@1 rose by 0.046 on \ds{tgbl-wiki} under exhaustive ranking and by 0.124 on \ds{tgbl-comment}. In list counts, the true destination moved to first place in about 22 more of the 484 \ds{tgbl-wiki} lists and about 372 more of the 3,000 \ds{tgbl-comment} lists. Hits@10 rose by 0.029 and 0.082. Pooled ROC-AUC rose by only 0.009 and 0.025. This secondary metric pools candidate rows across all lists, so it does not directly measure the within-list ranking task. At 1 hour on \ds{tgbl-wiki}, the Hits@10 and pooled-AUC intervals included zero, even though MRR and Hits@1 increased.

Under the original $C$ grid, the selected value was at a grid edge in 192 of 300 \ds{tgbl-wiki} logistic regression runs and 131 of 200 \ds{tgbl-comment} runs. The wider grid reduced these counts to 163 and 104. Every specified contrast kept its sign, and every contrast kept its interval status except the 1-hour shuffled control on \ds{tgbl-wiki} described above. No estimate changed by more than 0.008 MRR, and the \ds{tgbl-comment} estimates agreed within 0.0001 at every horizon.

\section{Discussion}

Later events improved the ranking of hidden pairs on both datasets, and the gain was largest where the past contained the least information about the targets. On \ds{tgbl-comment}, the 26-feature model raised MRR from 0.485 to 0.593 at 24 hours. The reciprocal of MRR corresponds to the harmonic mean of the ranks, and for \ds{tgbl-comment} this quantity decreased from approximately 2.06 under past-only information to 1.69 when 24 hours of later events were added, in lists of up to 101 candidates. Under exhaustive ranking on \ds{tgbl-wiki}, it decreased from approximately 1.67 to 1.57 among roughly 925 candidates. In relative terms, under the matched historical-random protocol, the gain is 22\% of the past-only MRR on \ds{tgbl-comment} (0.108\,/\,0.485) and 7\% on \ds{tgbl-wiki} (0.046\,/\,0.617). This difference is consistent with how much the past already reveals, since 71.7\% of \ds{tgbl-wiki} target pairs had appeared in the past, compared with only 10.7\% of \ds{tgbl-comment} target pairs. Because the two datasets also differ in their past input, with a 30-day lookback on \ds{tgbl-comment}, this difference cannot be attributed to how often pairs appear again alone.

For anyone repairing an incomplete interaction record, a practical question is how long to wait for later events. On \ds{tgbl-comment}, 6 hours of later events already gave 0.096 of the 0.108 gain reached at 24 hours, so a short window captured most of the benefit. On \ds{tgbl-wiki}, the gain still rose from 0.029 to 0.039 between 6 and 24 hours, so a longer window continued to help. The length of the later window should therefore be treated as a parameter chosen on validation episodes for each dataset, rather than fixed in advance.

The pair heuristic gives a second practical point. A later-window version of EdgeBank, which replaces the past pair count with the later pair count, lost 0.157 MRR under exhaustive ranking on \ds{tgbl-wiki}. The same later events, added alongside the past features in the 26-feature model, gained 0.039. Later evidence is therefore more useful when it is added to the past record than when it replaces it.

The two controls also show that the datasets respond differently to changes in the later record, although they do not identify a single mechanism. On \ds{tgbl-wiki}, removing the later events from query sources to historically observed destinations reduced the gain by 59\%. On \ds{tgbl-comment}, the same removal reduced it by only 10\%, while shuffling later destinations reduced it by 67\%. One possible reading is that the \ds{tgbl-comment} gain depends more on where other users' later activity goes, through destination activity and two-hop paths, than on the query source's own later events. Both controls change several properties at once, so this reading needs separate tests of the pair and two-hop features before it can be confirmed.

Finally, the choice of metric affects whether the gain is visible at all. At 24 hours on \ds{tgbl-wiki}, pooled ROC-AUC rose by 0.009, while Hits@1 rose by 0.046. A study that reported only pooled AUC could conclude that later events add little. Within-list metrics such as MRR and Hits@$k$ match the reconstruction task more closely, because the aim is to place the true pair near the top of each source's list.

\subsection{Limitations}
This study intentionally uses heuristics and logistic regression rather than temporal graph neural networks, so that the information given to each model is the only thing that changes between conditions. Whether learned memory models gain more or less from later events is therefore not tested here. The evaluation also uses three test episodes per dataset. Our intervals describe variation across users within these episodes and across five seeds, and the seeds share the same episodes, so they are not independent replications. Clustering the bootstrap by source user also does not account for dependence between lists that share destinations. More test episodes, especially on \ds{tgbl-comment}, are needed to assess whether the gains generalize to other time periods.

The gaps in this study are created by hiding events at chosen times. Real gaps may be longer, may coincide with outages that also change user behaviour, or may remove events unevenly, so the gains reported here may not carry over directly to every real record. The two datasets also differ in their past input and primary protocol. \ds{tgbl-wiki} uses its full past and exhaustive ranking, while \ds{tgbl-comment} uses a 30-day lookback, a cap of 1,000 targets per episode and historical-random candidates. We excluded targets whose destination never appeared in the past, so our results cover the historical universe rather than every pair active in the hidden window. Finally, we did not compute intervals for the control shares, and we did not test the pair and two-hop feature groups separately.

These limitations point to direct next steps. Beyond the separate feature tests and additional test episodes, a natural extension is to feed the recovered pairs back into the record and test whether they agree with the events observed after the gap, and whether they improve later forecasts.

\section{Conclusion}
Later events improved the ranking of directed pairs hidden in a two-hour window on both \ds{tgbl-wiki} and \ds{tgbl-comment}, with 24-hour MRR gains of 0.039 to 0.046 on \ds{tgbl-wiki} and 0.108 on \ds{tgbl-comment} in the main comparisons. These results suggest three practical points. First, retrospective reconstruction should be evaluated separately from forward prediction, with candidate lists kept fixed across information conditions. Second, later events should be added to past features rather than used in place of them, since replacing the past pair count lost up to 0.157 MRR on \ds{tgbl-wiki}. Third, the length of the later window should be chosen for each dataset, because 6 hours already gave 89\% of the 24-hour gain on \ds{tgbl-comment} but only 74\% on \ds{tgbl-wiki}.

\section*{Funding}
This research received no specific grant from any funding agency in the public, commercial or not-for-profit sectors.

\section*{Acknowledgements}
AI-based tools were used for language editing. The author takes full responsibility for the content of this manuscript.

\section*{Data availability}
The \ds{tgbl-wiki} and \ds{tgbl-comment} datasets are publicly available through the Temporal Graph Benchmark~\cite{huang2023tgb}. Episode definitions, candidate lists, feature hashes, model scores and evaluation outputs will be deposited with the code, and are available from the author on request until then.

\section*{Code availability}
The code for data preparation, feature construction, model fitting and evaluation, together with the protocol and amendment records, will be released in a public repository on publication.

\section*{Competing interests}
The author declares no competing interests.

\section*{Ethics}
The study uses publicly released, pseudonymized benchmark data. No human participants were recruited, and no new personal data were collected.

\section*{Author contributions}
Amarnath R conceived and designed the study, implemented the pipeline, ran the experiments, analyzed the results and wrote the manuscript.


\end{document}